\def\omegal{\omega_\ell}
\documentstyle[aps,prl,epsf]{revtex}
\begin{document}
\draft
\twocolumn[
\hsize\textwidth\columnwidth\hsize\csname@twocolumnfalse\endcsname 

\title{Low-frequency peak in the magnetoconductivity of a\\ non-degenerate
  2D electron liquid}

\author{Frank~Kuehnel$^1$, Leonid
  P.~Pryadko$^2$, and  M.I.~Dykman$^1$} 

\address{$^1$Department of Physics and Astronomy, Michigan State
  University, East Lansing, MI 48823\\ 
$^2$School of Natural Sciences, Institute for Advanced Study, Olden Lane,
  Princeton, NJ 08540}
\date\today
\maketitle
\widetext
\begin{abstract}
We study the frequency-dependent magnetoconductivity of a strongly
correlated nondegenerate 2D electron system in a quantizing magnetic
field.  We first restore the single-electron conductivity from
calculated 14 spectral moments.  It has a maximum for
$\omega\sim\gamma$ ($\hbar\gamma$ is the disorder-induced width of
the Landau level), and scales as a power of $\omega$ for $\omega\to
0$, with a universal exponent.  Even for strong coupling to
scatterers, the electron-electron interaction modifies the
conductivity for low and high frequencies, and gives rise to a
nonzero static conductivity.  We analyze the full many-electron
conductivity, and discuss the experiment.
\end{abstract}
]
\narrowtext

\begin{figure}[htbp]
    \epsfxsize=3.0in%\columnwidth%
    \epsfbox{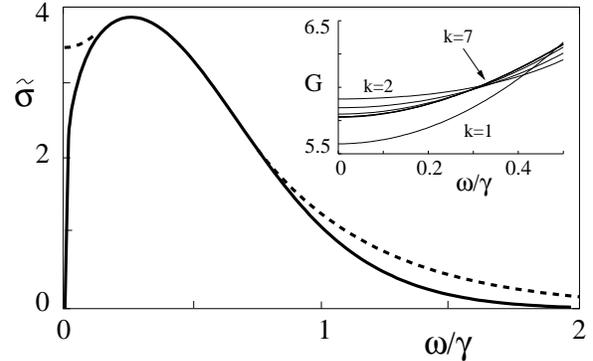} \caption{Reduced microwave
      con\-duc\-tiv\-i\-ty (\protect\ref{kubo_formula2}) of a
      non-inter\-act\-ing 2DES in a short-range disorder potential for
      $k_BT\gg \hbar\gamma$ (solid line).  The electron-electron
      interaction results in flattening of the conductivity for
      $\omega \protect\alt \omega_l$ (\protect\ref{cutoff}), and in a
      much slower decay of $\tilde\sigma$ for moderately big
      $\omega\gg \gamma$, as shown by dashed lines.  Inset:
      convergence of the interpolation factor $G$ (\protect\ref{G-function})
      with the increasing number of moments $2k$.}
    \label{fig:conduct}
\end{figure}

One of the most interesting problems in physics of low-dimensional
systems is the effect of the electron-electron interaction (EEI) on
electron transport.  In many cases the EEI is the major factor,
fractional quantum Hall effect (QHE) being an example.  At the same
time, single-electron picture is often also used for interpreting
transport, as in the integer QHE.  Another closely related example is
magnetotransport of a low-density two-dimensional electron system
(2DES) on helium surface\cite{Andrei-Book}.  For strong
quantizing magnetic fields, experimental data on electron transport in
this system are reasonably well described
\cite{Adams-88B,Heijden-88,Lea-97} by the single-electron
theory based on the self-consistent Born approximation
(SCBA)\cite{Ando-82}.  This theory does not take into account the
interference effects that lead to electron localization in the random
potential of scatterers.  Such a description appears to
contradict the phenomenology of the integer QHE, where all but
a finite number of single-particle states in the random potential are
{\it localized\/}\cite{Pruisken-localization,Huckestein-95}. The
static single-electron magnetoconductivity $\sigma_{xx}(0)$ must
vanish, as illustrated in Fig.~\ref{fig:conduct}, since the
statistical weight of the extended states is equal to zero.

In this paper we discuss the case where the EEI is strong and the
electrons are correlated, as for 2DES on helium and in fractional QHE.
Yet the characteristic force on an electron from the short-range
random potential may exceed the force from other electrons.  The
interrelation between the forces determines the effective strength of
the coupling to scatterers.  The analysis allows us to understand the
strong coupling limit and the crossover to weak coupling, and to
resolve the apparent contradiction between localization of
single-electron states and the experimental data for electrons on
helium.

We show that, for strong coupling to scatterers, the low-frequency
magnetoconductivity $\sigma_{xx}(\omega)$ of a nondegenerate 2DES
becomes {\it nonmonotonic\/}: it has a maximum at a finite frequency
$\omega_{\max}
\approx 0.3\gamma$, where $\hbar\gamma$ is the SCBA level broadening
[Fig.~\ref{fig:conduct}].  For small but not too small
$\omega/\gamma$, the conductivity scales as $\omega^{\mu}$ with a
universal exponent $\mu\approx 0.215$.  Whereas the onset of the peak
is a single-electron effect, the nonzero value of the static
conductivity and the form of $\sigma_{xx}(\omega)$ for big
$\omega/\gamma$ are determined entirely by the EEI.  We obtain an
estimate for $\sigma_{xx}(0)$ and analyze the overall shape of
$\sigma_{xx}(\omega)$ in the parameter range where
$\exp(\hbar\omega_c/k_BT)\gg 1$ and $k_BT\gg
\hbar\gamma$ ($\omega_c$ is the cyclotron frequency), the conditions
usually met in strong-field experiments on electrons on helium.

{\bf The single-electron conductivity} at low frequencies is
determined by the correlation function of the velocity of the guiding
center ${\bf R}$ of the electron cyclotron orbit in the potential of
scatterers. For $\omega\ll k_BT/\hbar$ and
$\exp(\hbar\omega_c/k_BT)\gg 1$, it can be written as 
$\sigma_{xx}(\omega) = (ne^2l^2\gamma/ 8k_BT)
\tilde\sigma(\omega),$ where $n$ is the electron density,
$l=(\hbar/m\omega_c)^{1/2}$ is the magnetic length, and 
$\tilde \sigma$ is the reduced conductivity,
\begin{eqnarray}
  \label{kubo_formula2}
  \tilde\sigma(\omega)=&& 
  -{2\hbar\gamma\over m\omega_c}\int_{-\infty}^{\infty}dt\,e^{i\omega t}
  \sum\nolimits_{{\bf q},{\bf q}'}\left({\bf q}\,{\bf q}'\right) \nonumber \\ 
  &&\times \bigl\langle \tilde V_{\bf q}\tilde
  V_{{\bf q}'}\exp\left[i{\bf
        q\,R}(t)\right]\exp\left[i{\bf q}'\,{\bf R}(0)\right]\bigr\rangle.
\end{eqnarray}
Here, $\langle\,\cdot\,\rangle$ stands for thermal averaging
followed by the averaging over realizations of the random potential
of defects $V({\bf r})$, and $\tilde V_{\bf q} = (V_{\bf
q}/\hbar\gamma)\,\exp(-l^2q^2/4)$ are proportional to the Fourier
components $V_{\bf q}$ of $V({\bf r})$.
We will assume that $V({\bf r})$ is Gaussian and delta-correlated,
\begin{equation}
  \label{potential}
  \langle V({\bf r})\,V({\bf r}')\rangle = v^2\delta({\bf
    r}-{\bf r}'),
\end{equation}
in which case $\hbar\gamma = (2/\pi)^{1/2}v/ l$\cite{Ando-82}.  

Time evolution of the guiding center ${\bf R}\equiv (X,Y)$ in
Eq.~(\ref{kubo_formula2}) is determined by the dynamics of a 1D
quantum particle with the generalized momentum and coordinate $X$ and
$Y$, and with the Hamiltonian
\begin{equation}
  \label{hamiltonian}
  H=\hbar\gamma \sum\nolimits_{\bf q} \tilde V_{\bf q}\,\exp(i{\bf
  qR}),\quad [X,Y]=-il^2.
\end{equation}

Because of the Landau level degeneracy in the absence of random
potential, the problem of dissipative conductivity is to some extent
similar to the problem of the absorption spectra of Jahn-Teller
centers in solids\cite{Stoneham-75}, which are often analyzed using
the method of spectral moments.  This method can be applied to the
conductivity (\ref{kubo_formula2}) as well\cite{Dykman-78}. It allows,
at least in principle, to restore $\sigma_{xx}(\omega)$. In addition,
the moments

\begin{equation}
  \label{MOM}
  M_k={1\over 2\pi\gamma}\int_{-\infty}^{\infty} d\omega
  \,(\omega/\gamma)^k\tilde\sigma(\omega)
\end{equation}
can be directly found from measured $\sigma_{xx}(\omega)$, and
therefore are of interest by themselves.

For $\omega, \gamma\ll k_BT/\hbar$, the states within the broadened
lowest Landau level are equally populated and the reduced conductivity
is symmetric, $\tilde\sigma(\omega) = \tilde\sigma(-\omega)$. Then odd
moments vanish, $M_{2k+1} =0$. For even moments, we obtain from
Eqs.~(\ref{kubo_formula2}), (\ref{MOM})
\begin{eqnarray}
  \label{general_MOM}
  M_{2k}=&&-2l^2
  \sum
  ({\bf q}_1\,{\bf q}_{2k+2})\,\bigl\langle\tilde V_{{\bf q}_1}\ldots 
  \tilde V_{{\bf q}_{2k+2}}\bigr\rangle\\
  &&\times \left[\left[\ldots\left[e^{i{\bf q}_1{\bf R}},\,
              e^{i{\bf q}_2{\bf R}}\right],\ldots\right],\,
      e^{i{\bf q}_{2k+1}{\bf R}}\right]\,e^{i{\bf q}_{2k+2}{\bf
  R}},\nonumber 
\end{eqnarray}
where
the sum is taken over all ${\bf q}_1,\ldots,{\bf q}_{2k+2}$.  The
commutators  (\ref{general_MOM}) 
can be evaluated recursively using 
\begin{equation}
  \label{commutator}
  \bigl[e^{i{\bf qR}},\,e^{i{\bf q}'{\bf R}}\bigr]=
      2i\sin\Bigl({1\over2}l^2{\bf 
      q}\wedge{\bf q}'\Bigr)\, e^{i({\bf q}+{\bf q}')\,{\bf R}}.
\end{equation}

{}From Eq.~(\ref{potential}), $\langle \tilde V_{\bf q}\tilde V_{{\bf
q}'}\rangle = (\pi l^2/2S)\exp(-l^2q^2/2)\delta_{{\bf q}+{\bf q}'}$, where
$S$ is the area. The evaluation of the $2k\,$th moment comes then to choosing
pairs $({\bf q}_i, -{\bf q}_i)$ and integrating over $k+1$
independent ${\bf q}_i$. From Eq.~(\ref{commutator}), the integrand is a
(weighted with ${\bf q}_1{\bf q}_{2k+2}$) exponential of the quadratic
form $(l^2/2)\sum {\bf q}_i\hat A_{ij}{\bf q}_j$, where $i,j = 1,\ldots,
k+1$. The matrix elements $\hat A_{ij}$ are themselves $2\times
2$ matrices, $\hat A_{ij} = -\hat I\delta_{ij} +
a_{ij}\hat\sigma_y$, where $\hat\sigma_y$ is the Pauli
matrix, and $a_{ij}=-a_{ji} = 0,\,\pm 1$. Because of the structure of
the matrices $\hat A$, the
moments $M_{2k}$ are given by {\it rational\/} numbers. For
$k=0,1,\ldots,7$ we obtain \cite{Kuehnel_to_be}

\begin{eqnarray}
  \label{m_values}
  M_{2k}=&&1;\, {3\over 8};\, {443 \over 1152};\, {25003\over 38400};\,
  {13608949709 \over 8941363200 };\,\nonumber\\ && 4.47809; \,15.7244;\,
  63.7499 
\end{eqnarray}
(we give approximate values of $M_{2k}$ for $k\geq 5$).

To restore the conductivity $\tilde\sigma(\omega)$ from the calculated
finite number of moments, we need its asymptotic form for
$\omega\gg\gamma$.  It can be found from the method of optimal
fluctuation\cite{Ioffe-81}, by calculating the thermal average in
Eq.~(\ref{kubo_formula2}) on the exact eigenstates $|n\rangle$ of the
lowest Landau band of the disordered system. All states $|n\rangle$
are equally populated for $k_BT\gg \hbar\gamma$. Their energies $E_n$
are symmetrically distributed around the band center ($E=0$), with the
density of states $\rho(E)\propto
\exp(-4E^2/\hbar^2\gamma^2)$\cite{Wegner-83}.  For large
$\omega/\gamma$, the conductivity is formed by transitions between
states $|n\rangle,\,|m\rangle$ with large and opposite in sign
energies $E_{n,m}$ ($|E_n-E_m|=\hbar\omega$). The major contribution
comes from $E_n=-E_m$. Only those configurations of $V({\bf r})$ are
significant, where the states
$|n\rangle,
\,|m\rangle$ are spatially close. However, the overlap matrix elements
affect only the prefactor in $\tilde\sigma$ \cite{Kuehnel_to_be}, 
and to logarithmic accuracy,
\begin{equation}
  \label{large_omega}
  \tilde\sigma(\omega) \propto [\rho(\hbar \omega/2)]^2\propto
  \exp(-2\omega^2/\gamma^2).
\end{equation}

Since the tail of the conductivity is Gaussian, one is tempted to
restore $\tilde\sigma(\omega)$ from the moments $M_n$ using a standard
expansion in Hermite polynomials, $\tilde\sigma(\gamma x)=\sum_n
c_n\,H_n(\sqrt 2 x)\exp(-2 x^2)$. From (\ref{MOM}), the coefficients
$c_n$ are recursively related to the moments $M_{k}$ with $k\leq n$.
However, for the moments values (\ref{m_values}), such an expansion
does not show convergence. This indicates possible {\it
  nonanalyticity\/} of the conductivity at $\omega=0$.

For $\omega\to 0$, the conductivity can be found from scaling
arguments\cite{Chalker-Coddington,Wang-Fisher-Girvin-Chalker} by 
noticing that it is formed by states within a narrow energy band
$|E|\ll \hbar\gamma$. The spatial extent of low-energy states is of
the order of the localization length $\xi\sim l\,|\varepsilon|^{-\nu}$,
where $\varepsilon = E/\hbar\gamma$ and $\nu=2.33\pm0.03$ is the
localization exponent\cite{Huckestein-95}.  The frequency $\omega$, on
the other hand, sets a ``transport'' length $L_\omega\sim
l\,(\gamma/\omega)^{1/2}$. It is the distance over which an electron
would diffuse in the random field $V({\bf r})$ over time $1/\omega$,
with a characteristic diffusion coefficient $D=l^2\gamma$, if there
were no interference effects. For large $\xi, L_\omega\gg l$, the
scaling parameter can be chosen as $g=(L_{\omega}/\xi)^{1/\nu}\sim
|\varepsilon|\,(\omega/\gamma)^{-1/2\nu}$
\cite{Wang-Fisher-Girvin-Chalker,Sondhi-unpublished}. 
The conductivity $\tilde\sigma(\omega)$ is determined by the states
within the energy band where $g\alt 1$. For high $T$, all these states
contribute nearly equally, and
\begin{equation}
  \label{eq:singular-conductivity}
  \tilde\sigma(\omega)\propto \omega^\mu \; (\omega\to 0),\;
  \; \mu=(2\nu)^{-1}\approx 0.215.
\end{equation}

With Eqs.~(\ref{large_omega}), (\ref{eq:singular-conductivity}), the
conductivity can be written as
\begin{equation}
\label{G-function}
\tilde\sigma(\omega)=x^{\mu}G(x)\exp(-2x^2),\;x=|\omega|/\gamma.
\end{equation}
The function $G(x)$ ($x \geq 0$) can be expanded in Laguerre
polynomials $L_n^{(\mu-1)/2}(2x^2)$, which are orthogonal for the
weighting factor in Eq.~(\ref{G-function}). 
We have restored the corresponding expansion coefficients from the
moments~(\ref{m_values}). 
The
resulting conductivity is shown in Fig.~\ref{fig:conduct} with solid
line. The expansion for $\tilde\sigma$ converges rapidly for $\mu$
between 0.19 and 0.28 (as illustrated in Fig.~\ref{fig:conduct} for
$\mu = 0.215$), 
whereas outside this region the convergence deteriorates.

{\bf The electron-electron interaction} (EEI) can strongly affect the
magnetoconductivity even for low electron densities, where the 2DES is
nondegenerate.  Of particular interest for both theory and experiment
are many-electron effects for densities and temperatures where
$\Gamma\equiv e^2(\pi n)^{1/2}/k_BT \gg 1$.  The 2DES is then {\it
strongly correlated\/} and forms a nondegenerate electron liquid or,
for $\Gamma > 130$\cite{Wigner-crystal}, a Wigner crystal. The motion
of an electron is  mostly thermal vibrations about the
(quasi)equilibrium position inside the ``cell'' formed by other
electrons.  For strong $B$, the characteristic vibration
frequency is $\Omega_p= 2\pi e^2n^{3/2}/m\omega_c, \, \Omega_p\ll
\omega_c$ (for a Wigner crystal, $\Omega_p$ is the zone-boundary
frequency of the lower phonon branch\cite{Andrei-Book}). We will assume that
$k_BT \gg \hbar\Omega_p$. Then the vibrations are quasiclassical, with
amplitude $\delta_{\rm fl} \sim (k_BT/e^2n^{3/2})^{1/2}\gg l$.  

The restoring force on an electron is determined by the electric field
${\bf E}_{\rm fl}$ from other electrons. The distribution of this
field is Gaussian, except for far tails, and
$\langle E^2_{\rm fl}\rangle
=F(\Gamma)\, n^{3/2}k_BT$, with $F(\Gamma)$ varying only slightly, from
$8.9$ to $10.5$, in the whole range $\Gamma \agt 20$
\cite{Fang-Yen-97}. Since  $\delta_{\rm fl}
\gg l$, the field ${\bf E}_{\rm fl}$ is uniform over the electron
wavelength $l$.  The electron motion can be thought of as a
semiclassical drift of an electron wave packet in the crossed fields
${\bf E}_{\rm fl}$ and ${\bf B}$, with velocity $c E_{\rm fl}/B$.

In the presence of defects, moving electrons will collide with
them. If the density of defects is small and their potential $V({\bf
r})$ is short-range [cf.\ Eq.~(\ref{potential})], the duration of a
collision is
\begin{equation}
  \label{t_e} 
t_e=l(B/c)\,\langle E_{\rm fl}^{-1}\rangle  \sim (\hbar/el)\,
n^{-3/4}(k_BT)^{-1/2},
\end{equation}
and the scattering cross-section is
$\propto\gamma^2$.  For $\gamma t_e
\ll 1$, electron-defect collisions occur independently and
successively in time.  This corresponds to weak coupling to the
defects, and allows one to use a single-electron type transport
theory, with the collision rate $\tau^{-1}$ calculated for the
electron velocity $cE_{\rm fl}/B$ determined by the EEI, $\tau^{-1}\sim
\gamma^2t_e$\cite{Dykman-79}.  The many-electron weak-coupling results
have been fully confirmed by experiments\cite{Lea-97,Lea-98}. 

For $\gamma t_e \gg 1$, collisions with defects ``overlap'' in time,
which corresponds to the strong coupling limit.  In this case, from
Eqs.~(\ref{potential}), (\ref{t_e}), the characteristic force on an
electron from the random field of defects $F_{\rm rf}=\hbar\gamma/l
\gg eE_{\rm fl}$.  One might expect therefore that the EEI does not
affect the conductivity, and the single-electron theory discussed
above would apply. It turns out, however, that this is {\em not\/} the
case for the low- and high-frequency conductivity.

As a result of the EEI, the energy of an electron in the potential of
defects $V({\bf r})$ is no longer conserved.  The motion of each
electron gives rise to modulation of energies of all other electrons.
The overall change of the Coulomb energy of the electron system over a
small time interval is given by $\sum_n e\,({\bf E}_n\, \delta {\bf
  r}_n)$, where $\delta {\bf r}_n$ is the displacement of the $n$th
electron due to the potential of defects, and ${\bf E}_n$ is the
electric field on the $n$th electron from other electrons.  Clearly,
${\bf E}_n$ and $\delta {\bf r}_n$ are statistically independent.
This allows us to relate the coefficient of energy diffusion of an
electron $D_{\epsilon}$ to the characteristic coefficient $D=\gamma
l^2$ of spatial diffusion in the potential $V({\bf r})$,
\begin{equation}
  \label{energy_diffusion} D_{\epsilon}= (e^2/2)\,\langle E_{\rm
  fl}^2\rangle\,D \sim \gamma (\hbar/t_e)^2.
\end{equation}

Energy diffusion eliminates electron localization which caused
vanishing of the single-electron static conductivity.
The low-frequency boundary $\omega_{l}$ of the range of applicability
of the single-electron approximation can be estimated from the
condition that the diffusion over the energy layer of width
$\sim\delta\varepsilon_l = (\omega_l/\gamma)^{\mu}$ [which forms the
single-electron conductivity (\ref{eq:singular-conductivity}) at
frequency $\omega_l\ll\gamma$] occurred over the time $1/\omega_l$.
For $\mu=1/(2\nu)$, this gives
\begin{equation}
  \label{cutoff}
  \omegal/\gamma =
  C_{\rm l}\,(\gamma t_e)^{-2\nu/(\nu +1)},\quad
  C_{\rm l}\sim 1.
\end{equation}
All states with energies
$|\varepsilon|\alt \delta\epsilon_l$  contribute to the
conductivity for frequencies $\omega < \omega_l$. Therefore the many-electron
conductivity may only weakly depend on $\omega$ for $\omega <
\omega_l$, as shown in Fig.~\ref{fig:conduct}, and the static conductivity 
\begin{equation}
\label{static_me}
\sigma_{xx}(0)\approx \sigma_{xx}(\omega_l)\sim
(ne^2\gamma l^2/k_BT)(\gamma t_e)^{-1/(\nu + 1)}.
\end{equation} 
We note that there is a
similarity between the EEI-induced energy diffusion, which we could
quantitatively characterize for a correlated nondegenerate
system, and the EEI-induced phase breaking in QHE
\cite{Fisher-90,Polyakov-93}. The cutoff frequency $\omegal$ 
can be loosely associated
with the reciprocal phase breaking time.
 
The EEI also changes  the {\em high-frequency tail\/} of $\sigma_{xx}(\omega)$
in the range $\omega\ll \omega_c$. In the many-electron
system, the tail is formed by processes in which a guiding center of
the electron cyclotron orbit shifts in the field ${\bf E}_{\rm fl}$
(by $\delta {\bf R}$). The energy $\hbar \omega$ goes into the change
of the potential energy of the electron system $e{\bf E}_{\rm
fl}\delta {\bf R}$, whereas the recoil momentum $\hbar \delta R/l^2$
goes to defects.  For large $\omega$, it is necessary to find optimal
$\delta {\bf R}$ and ${\bf E}_{\rm fl}$. For weak coupling to defects,
$\gamma t_e\ll 1$, the
correlator (\ref{kubo_formula2}) can be evaluated to the lowest order
in $\gamma$,  which gives

\begin{equation}
  \label{tail}
  \tilde\sigma(\omega) = 
  \gamma\omega t_e^2\exp\left[-(2/\pi)^{1/2}\omega t_e\right].
\end{equation}
The exponential tail (\ref{tail}) is determined by the characteristic
many-electron time (\ref{t_e}), and the exponent is just linear
in $\omega$. For larger $\omega$, the decay of $\tilde\sigma$ slows down to
$|\ln \tilde\sigma(\omega)|
\propto (\omega t_e)^{2/3}/[\ln(\omega/\gamma)]^{1/3}$,  provided
$n^{1/2}\delta_{\rm fl}(\omega t_e)^{1/3}\ll 1$
\cite{Kuehnel_to_be}. This asymptotics results from anomalous tunneling 
\cite{Thouless} due to multiple
scattering by defects. It also applies for strong coupling to defects,
$\gamma t_e \gg 1$, and replaces the much steeper single-electron
Gaussian asymptotics (\ref{large_omega}).  

We note that the overall frequency dependence of $\sigma_{xx}(\omega)$
is qualitatively different for strong and weak coupling to
scatterers. In the latter case, $\sigma_{xx}$ is {\it maximal\/} for
$\omega=0$ and decreases monotonously with the increasing $\omega$, in
contrast to the behavior of $\sigma_{xx}(\omega)$ in the
strong-coupling case shown in Fig.~\ref{fig:conduct}. Both $\gamma$
and $t_e$ increase with the magnetic field, and by varying magnetic
field, electron density, and temperature one can explore the crossover
between the limits of strong and weak coupling.

It is interesting that both the static and the high-frequency
conductivities are {\it many-electron\/} even for $\gamma t_e \gg 1$,
where the coupling to defects is strong. It follows from
Eq.~(\ref{static_me}) (see also Fig.~\ref{fig:conduct}) that the
many-electron $\sigma_{xx}(0)$ is of the order of the single-electron
SCBA conductivity $\sigma_{xx}^{\rm SCBA}(0) = (4/3\pi)ne^2\gamma l^2$
\cite{Heijden-88} for not extremely large $\gamma t_e$.  This is a
consequence of the very steep frequency dependence of the full
single-electron conductivity (\ref{eq:singular-conductivity}) for
$\omega\to 0$.

It follows from the above arguments that the random potential of
defects does not eliminate self-diffusion in 2DES for $\Gamma < 130$,
where the electrons form a nondegenerate liquid.  For electrons on
bulk helium, the results on the static conductivity apply also for
$\Gamma > 130$, where electrons form a Wigner crystal.  In this case
the random field comes from thermally excited ripplons or (for $T \agt
1$~K) from helium vapor atoms\cite{Andrei-Book}.  Ripplons, although
they are extremely slow, do not pin the Wigner crystal
\cite{Dykman-81} (we note that, for  scattering by ripplons, 
$\gamma\propto T^{1/2}$).  Random potential of vapor atoms is
time-dependent (and also non-pinning). Vapor atoms stay within the
electron layer only for a time $t_v = a_B/v_T$, where $a_B$ is the
layer thickness and $v_T$ is the thermal velocity of the atoms.  For
strong magnetic fields one can have $\gamma t_v\gg 1$, and then if
$\gamma t_e\gg 1$, coupling to the vapor atoms is strong, as observed
in Refs.~\cite{Adams-88B,Heijden-88}.  The presented strong-coupling
theory describes the conductivity for arbitrary $t_e/t_v$ provided the
low-frequency cutoff of the single-electron theory $\omegal $
(\ref{cutoff}) is replaced by min($t_v^{-1}, \omegal $).

In conclusion, we have analyzed the magnetoconductivity of a
nondegenerate 2D electron liquid in quantizing magnetic field. This is
a simple and well-studied experimentally strongly correlated system,
where effects of the electron-electron interaction on transport can be
characterized qualitatively and quantitatively. It follows from our
results that, whereas for weak coupling to short-range scatterers the
conductivity $\sigma_{xx}(\omega)$ monotonically decays with
increasing $\omega$ ($\omega\ll
\omega_c$), for strong coupling it becomes nonmonotonic.
Even for strong coupling, the static conductivity is determined by
many-electron effects, through energy diffusion. It is described in
terms of the critical exponents known from the scaling theory of the
QHE. The frequency dispersion of $\sigma_{xx}$ disappears for
$\omega\alt \omegal\propto T^{\nu/(\nu +1)}$, for
temperature-independent disorder. In a certain range of magnetic
fields and electron densities, the value of $\sigma_{xx}(0)$
(\ref{static_me}) is reasonably close numerically to the result of the
self-consistent Born approximation, which provides an insight into
numerous experimental observations for electrons on helium surface.

We are grateful to M.~M.~Fogler and S.~L.~Sondhi for useful
discussions.  Work at MSU was supported in part by the Center for
Fundamental Materials Research and by the NSF through Grant no.
PHY-972205.  L.P.~was supported in part by DOE grant DE-FG02-90ER40542

\end{document}